# Computer Games Are Serious Business and so is their Quality: Particularities of Software Testing in Game Development from the Perspective of Practitioners

Ronnie E. S. Santos
Centro de Informática – Universidade Federal de Pernambuco
Brazil
ress@cin.ufpe.br

Cleyton V. C. Magalhães
Centro de Informática – Universidade Federal de Pernambuco
Brazil
cvcm@cin.ufpe.br

Luiz Fernando Capretz
Electrical & Computer Eng. Dept.
Western University
Canada
lcapretz@uwo.ca

Jorge S. Correia-Neto
Universidade Federal Rural de Pernambuco
Brazil
jorgecorreianeto@gmail.com

Fabio Q. B. da Silva
Centro de Informática – Universidade Federal de Pernambuco
Brazil
fabio@cin.ufpe.br

Abdelrahman Saher
Game Developer – Appsinnovate, Apps and Game Development
Cairo, Egypt
admin@a-saher.com

## ABSTRACT

**Context**. Over the last several decades, computer games started to have a significant impact on society. However, although a computer game is a type of software, the process to conceptualize, produce and deliver a game could involve unusual features. In software testing, for instance, studies demonstrated the hesitance of professionals to use automated testing techniques with games, due to the constant changes in requirements and design, and pointed out the need for creating testing tools that take into account the flexibility required for the game development process. **Goal**. This study aims to improve the current body of knowledge regarding software testing in game development and point out the existing particularities observed in software testing considering the development of a computer game. **Method**. A mixed-method approach based on a case study and an opinion survey was applied to collect quantitative and qualitative data from software professionals regarding the particularities of software testing in game development. **Results**. We analyzed over 70 messages posted on three well-established network of question-and-answer communities related to software engineering, software testing and game development and received answers of 38 professionals discussing differences between testing a computer game and a general software, and identified important aspects to be observed by practitioners in the process of planning, performing and reporting tests in this context. **Conclusion**. Considering computer games, software testing must focus not only on the common aspects of a general software, but also, track and investigate issues that could be related to game balance, game physics and entertainment related-aspects to guarantee the quality of computer games and a successful testing process.

## 1 INTRODUCTION

Over the last decades, computer games began to have a significant impact on society, replacing most of the traditional games and influencing how people in general spend their time. This increase has been boosted by the availability of new consoles, platforms and technologies, which have transformed the development and delivery of games as a continuous growth activity [1]. The proof of the popularity of games is evident in the number of successful games over time and in studies demonstrating that the software gaming industry has grown enormously, acquiring billions of dollars over the years, and reaching a well-established status along with other popular entertainment industries, such as music and cinema [2][3].

Although a computer game is a type of software, the process to conceptualize, produce and deliver a game can involve unusual features, such as the drive for novelty factors, creativity and artistic expression [4]. In fact, the differences between general software and a computer game, along with the differences in the development process to obtain both, has been discussed and published in previous studies.

For instance, Murphy-Hill, Zimmermann and Nagappan [5] identified that it is common in game development to work with less clear requirements, and creativity and the ability to communicate with non-engineers tends to be more highly valued. Further, their research demonstrates the hesitance of professionals to use automated testing techniques in this context, due to the constant changes in game requirements and design, and points out the need for creating testing tools that take into account the flexibility required by the game development process in defining tests.

Regarding software testing and the game development process, the literature presents studies that discuss different issues related to this theme. For instance, Lewis, Whitehead and Wardrip-Fruin [6] presented a taxonomy of possible failures identified in games to help practitioners to expose bugs in the game. Buhl and Gareeboo [7] described and discussed the use of automated testing to improve game development. Kasurinen and Smolander [8] analyzed how game development companies test their products, and identified that they tend to focus on values such as game content or user experience, instead of reliability or efficiency of the product. Washburn Jr. et al. [9] analyzed the posts of professionals published on a game web forum and used these to identify positive and negative characteristics of game development, including software testing, based on the experiences of developers. We discuss all of this evidence more fully in Section II.B.

This study aims to improve the current body of knowledge regarding software testing in game development, assuming that the development of a computer game has particular characteristics that could differ in comparison with the development of general software, as discussed in [4][5], by performing a mixed-method study to answer the following research question:

*RQ. What are the existing particularities in software testing, regarding the development of computer games?*

To answer this question we first analyzed, as a case study, a set of discussions posted on three well-established network of question-and-answer communities related to software engineering, software testing and game development. Following this, we developed a survey questionnaire and collected opinions from professionals working with software testing and game development to further explore the particularities of software testing in this context.

This paper is organized as follows. In Section 2 we present the conceptual background that supports this study. In Section 3 we describe the research method, instruments and techniques applied to answer our research question. In section 4 we present the main findings, which are discussed in Section 5. Finally, in Section 6, we present our conclusions and directions for future research.

## 2 BACKGROUND

This section presents the theoretical background that supports this study, as well as related research in a similar context.

### 2.1 Software Testing

Software testing is the dynamic verification and validation of software, in order to confirm that the product provides the expected planned behaviors. Over the years, the perceptions of software testing among practitioners and researchers has matured into a constructive view, since this activity is no longer seen as a phase that starts only after coding the software aiming to detect failures. Nowadays, software testing is, or should be, a pervasive set of activities throughout the entire software development and maintenance life cycle [23]. Similar to the software-development life cycle, a software-testing life cycle frequently divides the testing activities into five phases [24]:

- *Requirement Analysis*: focused on the understanding of requirements in terms of what will be developed and tested;
- *Test Planning*: focused on the construction of the artifacts that will guide the software development and execution: the test strategy and the test plan, which includes activities of estimations, selection of testing approaches, preparation of documents, definitions of tools and assignment of responsibilities;
- *Test Case Development*: focused on the process of writing test cases, and if required, the creation of scripts for automation. Also, can include the creation of test data;
- *Test Execution*: in summary, is the process of setup the environment required to execute the tests and then, perform all the tests (manual and automated), which also includes reporting test results, logging defects, verifications and retesting;
- *Test Closure*: discussions about the testing artifacts and evaluation of the process applied occur at this phase.

In summary, software testing characterizes an important part of the software development, representing, for software industry, one of the keys to reducing errors, maintenance and overall software costs [25]. Moreover, software testing is a dynamic activity, and the type of software under development can influence how this process is performed.

### 2.2 Related Studies on Software Testing and Game Development

Software testing is a very important phase of the game development process, and usually game testing is different from software testing in general, because the fast evolution of games has increased the complexity of this activity, which is requiring more diverse tests and simulations [2]. Over the years, researchers have pointed out this difference in software testing in the relation to games, and noted that although software quality is important in both games and in other types of software, the practice of testing appears to differ significantly [5].

A significant difference of the game testing process may be related to test automation. The literature suggests that game testing tends to be a more human-centered activity, since there is a certain level of difficulty in separating the user interface from the rest of the game [5]. In addition, human behavior is also an important factor that is present in computer games and this is difficult to automate. Despite its particular complexity when associated with game development, automated testing is a factor that could significantly improve the development of a game in terms of time and costs [7]. However, there is a lack of material and tools designed to facilitate test automation in games [5].

Game development seems to be friendlier to changes than general software development, even considering agile environments. In game development last-minute changes are usually expected and allowed, and this characteristic can affect even major changes [3]. Therefore, software testing tends to be more flexible in relation to game development, including aspects related to requirements, plans, test cases and estimations.

Finally, due to the complexity and variant characteristics of different types of games, reporting a failure in a game shall be different from reporting a failure in other types of software [6]. For instance, problems in games can be related to not only coding errors or designing mistakes, but also to balance (game rules), real time event occurrences, object boundaries, and many other factors.

In conclusion, over the years, researchers have gathered evidence on the differences between software testing in general and software testing in the context of game development. However, there is still a lack of information regarding this topic and some phases of the software-testing life cycle in games need to be further explored and discussed.

## 3  METHOD

In this study, a mixed-method approach was applied to collect quantitative and qualitative data from software professionals regarding the particularities of software testing in game development. In summary, the study was developed in two different stages:

*a)*  *Stage I*: characterized as a case study developed in two parts, to collect qualitative data posted in three communities of Stack Overflow, a reliable online community for software professionals to exchange information regarding several topics of software development. This type of environment have being recognized over the years for provide a large number of high quality useful answers regarding several topics, since their highly active users are typically experts in the main topic discussed in the community [26];

*b)*  *Stage II*: defined as an opinion survey, in which a questionnaire was designed to collect the opinions of professionals regarding testing processes in game development.

Figure 1 illustrates the methodological design followed in this study.

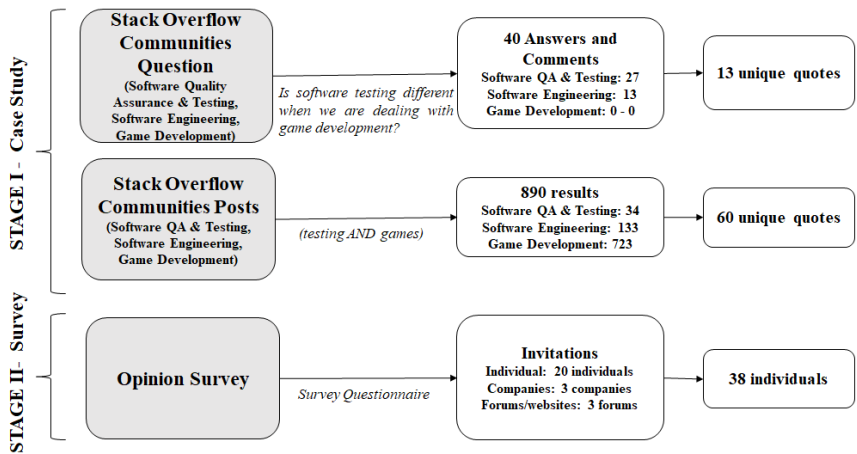

**Figure 1: Research Design.**

Hereafter, all methodological steps performed in this study are presented in detail.

### 3.1  Stage I: Case Study

A case study is an empirical inquiry that investigates a contemporary phenomenon within its real-life context, through detailed contextual analysis, when the boundaries between phenomenon and context are not evident [10] [11]. Case studies have been used in diverse research areas, and in software engineering this method may be suitable for exploring the complex interactions among people and technologies. Therefore, some of the activities and issues related to software development are better investigated in their natural settings, in order to achieve deeper understanding and improved results [12].

In this study, the guidelines for conducting case studies [12] and qualitative analysis [13] in software engineering were followed. In addition, general guidelines to perform case studies and qualitative analysis [14] [15] supported this process. Thus, we developed the following steps to perform this case study:

*1)  Getting Started*

The general motivation to perform this research comes from a discussion raised in a lecture about software testing performed in a software engineering class of a game development course. It was observed that there are differences when testing a game and a regular software (non-game) specially related to the amount of human interaction and different types of users in this context.

*2)  Selecting the Case*

Usually, a case study would require a real-life environment to collect data. Therefore, to complete the research, it would be necessary to identify and select a game development company interested in participating in this study. However, there were some difficulties in identifying such a specific company willing to take part of this study. Therefore, to overcome this limitation, the Stack Exchange Q&A platform was selected as a data source to collect information about the theme under study.

Stack Exchange Q&A is a network of communities that support discussions of users, experts and researchers on different and specific topics, such as computer programming, software engineering, and several other themes not related to computers or informatics [16]. This platform is a reliable source of data, since the community system of badges and reputation rewards users who provide high quality and well-researched answers. For this specific characteristic, the data stored in this network was used in several studies over the years [16] [17] [18] [19].

*3) Data Collection*

Data collection occurred in two different stages, representing two phases. In both phases, we collected data from three different online communities related to the general research question of this research: Software QA & Testing (https://sqa.stackexchange.com), Software Engineering (https://softwareengineering.stackexchange.com) and Game Development (https://gamedev.stackexchange.com).

First, we posted a question in all three communities, asking the opinion of users about the general differences between software testing and game testing. Those interested in discussing this topic were invited to answer the following question: *Is software testing different when we are dealing with game development?* At the end of this process, two weeks after the question was posted, 40 different answers and comments were collected, presenting the following distribution: a) Software QA & Testing: 27 results, b) Software Engineering: 13 results, c) Game Development: 0 results.

Internal rules for the Game Development community barred the question posted in the forum, because users affirmed that such question was discussing a broad topic, and therefore, not allowed in the community. Finally, after the filtering process, 13 different quotes were selected to be analyzed in the next step of the study. The filtering process eliminated answers indicating only links suggestions, posts commenting rules of the community and narrow opinions, just agreeing or disagreeing with answers but without presenting any significant clarifications.

Further, to access more evidence available in the communities, we collected an amount of data from previous existing posts in the platform. Thus, in each above cited community, a search using the term "testing AND game" was performed, retrieving an amount of 890 results among answers and comments to questions about software testing and game development, presenting the following distribution: a) Software QA & Testing: 34 results; b) Software Engineering: 133 results; c) Game Development: 723 results.

After a filtering process, we selected 60 different quotes referring to the theme to analyze in the next step of the case study. The difference between the total number of results and the number of quotations selected is related to the fact that many users used the term game as a synonym or slang for several activities when discussing software processes, such as planning game or discussions to clients. In addition, there were posts discussing the use of games to study and teach software engineering and software testing. Further, regarding the game development community, many posts referred to the process of testing tools or engines used to develop games instead of discussing the process of testing games under development.

*4) Data Analysis*

Qualitative data analysis was applied in this case study, due to the nature of data previously collected - textual data extracted from the discussions obtained from the three communities. Following the guidelines, a qualitative analysis aims to consolidate, reduce, and interpret data obtained from various sources, and make sense of them [15]. This process involves labeling and coding all data in order to categorize, and synthesize information [13].

Data analysis started with open coding of the collected quotations, followed by the construction of post-formed codes referring to a particular piece of text (Figure 2). Following the guidelines, we constantly compared the codes that emerged and then grouped them into categories (Figure 3). As the process of data analysis progressed, we described each identified category as a particularity of software testing in game development and related to one specific phase of the software-testing life cycle following the definitions presented on SWEBOK [23].

**Selected quotation**
"(...) usually that involves automating test cases, but as games are graphics intensive, you can't automate screens and animations as tests, but your game logic could be automated."

**Key point**
"can't automate screens and animations as tests."

**Code**
Difficulty with automation

**Figure 2: Building Codes**

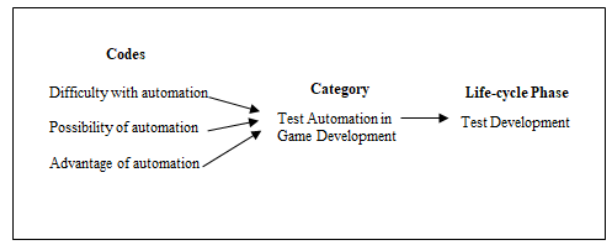

**Figure 3: Building Categories**

*5) Reaching Closure*

Following the guidelines, we compared and contrasted the results obtained in data analysis with findings from the literature in order to raise generalizability at the theoretical level. Finally, a qualitative case study usually requires a member checking phase to consolidate the results, and improve accuracy, credibility, and internal validity of the process [20]. However, the nature of this case, a network of question-and-answer communities, made impracticable the process of re-contacting users to discuss opinions posted months or years before this study. Further, the results obtained in this stage of the study were obtained through an interpretative perspective of data collected from a specific context; therefore, generalizations of results can be impracticable. However, the results can be re-analyzed to verify transferability to specific contexts.

### 3.2 Stage II: Survey

By the end of stage I, the findings obtained from the case study revealed a set of important particularities observed in the process of software testing in the context of games development. However, there was not enough evidence identified to cover all the phases of the software-testing life cycle. Therefore, the second stage of this study aimed to collect further information about the topic, applying a different approach, an opinion survey.

A survey can be defined as a type of research in which individuals are invited to answer questions about one topic or phenomenon and the information provided is used to discuss a topic under study. Thus, in this stage, the guidelines of Kitchenhan and Pfleeger [21] and Linaker et al. [22] were followed to perform a cross-sectional survey, and a questionnaire was applied to collected data from a selected sample of participants. The steps performed in this second stage of the study are described below.

*1) Setting Objectives and Designing the Survey Questionnaire*

In this survey the general objective was to collect opinions from different types of professionals regarding the differences among the testing process in general software development and game development. Therefore, following the guidelines, an instrument was developed for a team composed of experts with both research and domain expertise, to provide both technical and practical knowledge about the topic under investigation [22].

In this sense, the questionnaire was constructed by two researchers, with previous experience as software testers in industry and teaching experience in a Game Development course. Further, two academic researchers (PhD professors) reviewed the questionnaire. Both researchers have experience in themes related to software engineering and empirical studies.

To elicit opinions from professionals, the questionnaire included both closed and open questions. The general idea was the design of an instrument that could collect descriptive information from participants and their experience with software testing and with games, along with opinions regarding the information already collected and analyzed in the literature and the case study, and finally, information regarding evidence not observed in the case study. Below, the description of each part of the questionnaire is presented.

- Demographic Questions: Questions designed to collect descriptive information that could characterize the participants of this study;
- Open Question I: Questions designed to collect broad qualitative data about the characteristics of software testing in game development;
- Open Question II: Questions designed to collect qualitative data and that could assess information about the particularities of software testing in game development regarding each specific phase of the software-testing life cycle.

As recommended in the guidelines, a pilot questionnaire was tested and validated in order to identify problems with the questionnaire and responses. After validation and adjustments based on the considerations received from three specialists, two versions of the final questionnaire were implemented and distributed in English and Portuguese. Below we present the English version of the questionnaire.

**Table 1: Survey Questionnaire**

| Groups | Questions |
|---|---|
| **Demographic Questions – Descriptive Information** | Q1 – Name (Optional) |
| | Q2 – Email (Optional) |
| | Q3 - Organization you work for (Optional) |
| | Q4 - Highest completed level of education |
| | Q5 - Years of professional experience in Software Development (in years) |
| | Q6 - Current job position/ role |
| | Q7 - Years of experience in your current position/ role |
| | Q8 - Years of experience working with game development |
| | Q9 - How frequently would you say that you interact with games? ( ) Never ( ) Rarely ( ) Sometimes ( ) Often ( ) Almost Always |
| | Q10. Do you have any experience with software testing automation? |
| **Open Questions I – General Opinions regarding testing games** | Q11. In your opinion, how is software testing different when the software being developed is a game? |
| | Q12. What aspects of software testing would you consider as different or particular to the process of testing a game? |
| **Open Questions II – General Opinions regarding software-testing life cycle in game development** | *Requirement Analysis* |
| | Q13. What would you say is different when a testing team is working on understanding the requirements of a game? |
| | Q14. Would you agree that "a game development process seems to be more friendly to changes in requirements than the development of software in general"? |
| | *Test Planning* |
| | Q15. Test planning is the phase of software testing commonly related to the definition of a test strategy and a test plan, which includes activities of estimation, selection of testing approaches, definitions of tools and assignment of responsibilities. What particular characteristics would you say that these tasks have when a game is under development? |
| | *Test Case Development* |
| | Q16. Two important steps of software testing are the process of writing test cases and creating scripts to automate the execution of tests. Do you believe that there is any relevant difference in performing these tasks when a game is the object of the test? |
| | Q17. What would you say is most important to observe and consider when developing test cases for games? |

| |
|---|
| Q18. What would you point to as relatively harder to automate when testing games? |
| *Test Execution* |
| Q19. In summary, testing execution is the process of test and retest features, and then, report test results. How might this process vary or differ when the software under development is a game? |
| Q20. Do you believe there are types of bugs that are present only or mostly in games? Which would be they? |
| *Test Closure* |
| Q21. Test closure is the last phase of the software-testing life cycle and is related to evaluation of the whole testing process. How would you define a successful testing process of a game and how is that different from testing used for regular software |

*2) Population, Sample and Procedure*

Similar to the case study, this survey also faced some limitations regarding the selection of participants to answer the questionnaire. Firstly, the guidelines suggest that a research survey needs a well-defined target population, which means a group of individuals to whom the survey applies, that is, the total number of individuals who are able to answer the questionnaire. Therefore, a target population is represented as a finite list of all its members.

However, in this study, a broad topic is under investigation and the exact population is undefined, because considering the main research problem "software testing in game development," a well-defined population would be composed of all software testing professionals working with game development in the world. Thus, as discussed in the guidelines and similar to the results of the case study, the results of this survey cannot be universally generalizable for a population in a positivist perspective. However, at the theoretical level, the results can support analytical generalization and potential transferability to other contexts.

Following the guidelines, we performed the sampling process applying a convenience sampling. In this process, we obtained responses from professionals who were available and willing to take part in the study and individuals from the personal contacts list of the authors that had the profile to participate in the study were personally invited to answer the questionnaire.

In this case, the ideal profile of participants would be professionals working directly in testing activities in game development companies or game development projects, such as testers, developer-testers, QAs, test managers and test leaders. However, similar to the previous stage (case study), there was a limitation in identifying and contacting a representative number of professionals with this specific characteristic to participate in this study.

To overcome this problem, we extended the sample to include professionals with other characteristics, and included individuals who had enough previous experience to opine about the theme under investigation. This extension included general software professionals, such as developers, testing professionals, managers and software analysts from different types of software companies, but who have worked with games during any period of their professional life. For that, the survey included a question to investigate whether the respondent had experience working with game development at any time or not. Thus, we grouped the individuals invited to answer the questionnaire into two groups.

- Group I: We invited professionals working with software testing in three companies based in Brazil to answer the survey. These companies are characterized as follows: Company A is a test center that holds a partnership with an international mobile phone company. Company B also holds a partnership with an international mobile phone company, and the agreement includes not only testing activities, but also the development of new products. Company C is a private software organization specialized in software development and innovative software solutions in several business domains, such as finance, telecommunications, government, industry, services, and energy. All three companies had in their portfolio the development and/or testing of games and apps based on games. In this process, the invitation, along with the survey questionnaire (Portuguese version), was emailed to project managers and team leaders and they were asked to forward the invitation to their teams;
- Group II: We selected professionals working with software testing and with game development from the contact list of the authors and invited them to answer the questionnaire. This group also included individuals identified on LinkedIn as interested in participating of this research. In this case, the invitations were directly sent by email to authors' contacts, sent by direct message to individuals identified on LinkedIn, and also posted on the authors' public pages.

*3) Data Analysis*

By the end of the data collection, 41y different professionals answered the survey questionnaire providing opinions on their experience about game software testing. Then, we applied both qualitative and quantitative analysis to explore the content of the collected opinions.

To analyze the textual data collected from open questions, we applied the process that involves labelling and coding the quotations provided by the respondents. This is the same process applied on the case study and described in Fig. 2 and Fig. 3.

On the other hand, the answers of closed questions were analyzed using descriptive statistics in order to present the characterization of the sample and the distribution of participants' answers. In this phase, we explored the data with support of MS Excel™, which we also used to generate graphics and tables.

## 4 RESULTS

This section presents the results obtained from this research. In summary, each stage and step performed produced data input to the next stage, and at the end, the whole study presents a set of findings regarding the particularities of software testing in game development.

### 4.1 Case Study Results

The general results pointed out that testing a game might differ from testing a non-game software in many aspects. Testing a game is a process that usually face some particularities related to coverage of the tests regarding the physics of the game and, the variety of possible test scenarios and flows. Further, estimations regarding dynamic and possible variable requirements, together with metrics that are not easily measured, such as entertainment or fun, are among the factors that should be considered by software testers in this context.

### 4.1.1 Case Study - Part I

The case study started by asking individuals of three different online communities on Stack Exchange (Software QA & Testing, Software Engineering and Game Development) whether software testing is different when they are dealing with game development or not. Following this, while individuals commented their experience working in game development, we collected 40 answers with 13 unique quotes regarding software testing, as illustrated in Figure 1. This data was analyzed following the qualitative process described in Figure 2 and 3, and the results obtained demonstrated that the main difference in testing games is related to the phases of test planning and test case development.

Considering Test Planning, individuals discussed that, in games, there are specific measures that are uncommon or absent in regular software, which are difficult to measure or even observe due to their abstract nature. These measures are related to fun, entertainment, gameplay and other user experience aspects and their occurrence on the game could be too complex to plan and evaluate in terms of software testing. Data analysis showed that practitioners see on these factors one of the greatest challenges of game testing.

*"It's difficult for the testing tool to measure the degree of entertainment or the consistency and realism of the scenes the user see". (IN012)*

*"For instance 'fun factor' testing is something unique to games. Since they are an entertainment product. Games are not only supposed to work intuitively and it should provide a good user experience." (IN008)*

Further, an effective test planning process for games should consider specific *professional skills* for testers that will be allocated to develop and execute the tests. These professionals should be able to understand the principles and the characteristics behind games and especially understand the game development context, to guarantee the quality of the game and a successful testing execution, as demonstrated in the following quotes.

*"Game tester should have the same general knowledge base as a software tester but with a special focus of what makes games unique". (IN007)*

*"I've never found a more dedicated group of testers in any other domain, since they want to test the software. They're having fun. They're addicted and sleeping next to the computer". (IN013)*

Regarding Test Case Development, individuals pointed out that the *test cases creation* is an activity that can involve a more complex process in testing games, due to the enormous alternative ways that players can execute the software, differently from individuals using a software with a well-defined number of actions, as illustrated in the quotations below.

*"For regular software, it can be assumed that most (legitimate) users will attempt to use the software as designed, attempting to find the happy path. With game testing, players will often attempt to break some aspects." (IND002)*

*"Games are more immersive and interactive than other software, gamers will try almost everything beyond your imagination." (IN004)*

*"Simply put, the number of possible unique ways to do something in context of games, can be mathematically, very very large." (IN010)*

Software test automation is another activity that requires attention when the software under testing is a game due to the characteristics of game requirements, which might be not just associated to high levels of changes, but also, related to multifaceted elements of user interaction and experience, turning many games' features very difficult to develop test automation, as presented below.

*"You can auto-test things like file format loaders but how will you write a unit test that taking damage from a bomb exploding while simultaneously trying to grab the bomb and put your shield." (IN001)*

*"Test automation is used to mainly test simpler, non-interactive game aspects, such as making sure there is no gap on this map, all trees are taller than 3 meters." (IN005)*

To this point, this study gathered a relatively small number of particularities regarding game testing, by asking practitioners to spontaneously comment their experiences in game development. In this process, we collected evidences about only two phases of the software-testing life cycle, so far. Therefore, the second stage of this case study selected and analyzed about 890 messages posted on the three online communities of Stack Exchange, identifying new evidence on 60 unique quotes. These messages were posted on the communities over the years, and the first identified quote related to game testing is from 2010.

### 4.1.2 Case Study - Part II

From the amount of 60 messages and comments, over 73% of messages were posted on the Game Development community, followed by 13% of messages found in the Software Testing community and more 13% in the Software Engineering community. This new amount evidence was successfully applied to enlarge the results of this study, since the data collected brought new information regarding other phases of the software-testing life cycle in games, such as Requirement Analysis. However, over 91% of messages were related to Test Planning and Test Case Development, which was applied to confirm the findings obtained in the first phase of this case study and to improve the understanding acquired so far, adding new information to the research.

Regarding Requirements Analysis, individuals described game features as changeable and naturally variable, since the list of requirements are used to evolve and change over time as designers and developers identify better approaches to improve user interaction and experience in the game, as illustrated by the following quotes.

"Games rarely have exact specifications when started. And if they do, they always change and evolve during the development process." (IN043)

"I'll have to make changes to already released episodes, requiring me to keep testing them." (IN071)

Further, as demonstrated in the quotes below, the way to improve the quality of requirements and consequently the tests themselves would be related to how close users and stakeholders are involved in the Requirements Analysis process to help testers to define what they expect from the game.

"One shouldn't start by writing tests or code, but instead, should get back to the stakeholders and work with them to produce sane requirements." (IN026)

"Requirements validation could include things like unit testing and verification of features with users, which is invaluable for a developer without a large QA department at his back." (IN062)

Regarding Test Planning, findings from the second stage of the case study confirmed particularities observed before, such as the need for software testers with skills of a gamer in some level, which will facilitate the process of identifying and reporting bugs, as illustrated below.

"Games are more reliable if at least one member of the test group is a skilled gorilla tester, masterful at ad-hoc testing." (IN034)

"Game companies need testers who are genuinely interested in what they're doing." (IN034)

"You need professional testers who thoroughly test all the edge-cases of your game, systematically looks for bugs." (IN034)

Further, aspects related to metrics that are difficult to observe and access in games were also observed by individuals at this point.

"There is no test for fun in games, and there is no test for usability of a graphical user interface." (IN042)

"Large part of the bugs are art/graphics related (holes in collision meshes, wrong textures whatever, glitch in the depth of field shader)." (IN047)

The general idea to overcome this problem could be associated to the strategy applied in the testing game process. Multiple approaches combining unit tests, manual testing, automated tests and exploratory tests are not only appreciated in this context. A test strategy, gathering all this approaches, might be mandatory to obtain the level of quality expected in a game. Moreover, despite the general principle that exhaustive testing is impractical in software development, game testing needs to predict certain level of test repetitions (higher than in regular software) and the associated costs to that.

"Combining multiple approaches to achieve a high level of confidence in the functional quality and reliability of your game." (IN035)

"While in some cases you can unit test everything, it's usually not practical to achieve 100% coverage and, especially in games, can be quite difficult." (IN037)

"Spending hundreds of hours developing test automation for every little nook and cranny can also be bad. Find a middle ground between automation and playtesting." (IN040)

"What no unit test can do, however, are the complex interactions of multiple paths of game logic interacting." (IN055)

"In game testing, you may be asked to do things like play the same level 800 times until you can figure out the exact steps needed." (IN060)

For Test Case Development, the new findings confirmed the complex process of writing test cases for games due to the enormous number of scenarios and flows that needs to be checked in the game.

"With too many modes and flags, the game can quickly become very difficult to test, because of the number of possible variants." (IN024)

"For a single action game, it may take dozens, or even hundreds of times to playtest each level to make sure they are balanced." (IN056)

Nevertheless, in this second stage of the case study, by exploring messages and comments posted over the years, it was possible to identify evidence proposing solutions to overcome the problem related to too many scenarios, flows and different types of users that have to be considered, turning the activity of writing test cases into a less complex process. Thus, for games, this activity might include the definition of personas that represent groups of players divided by type, involve the designers and analysts in the process of defining test cases, in order to prioritize flows and identify lack of coverage, and finally, using exploratory testing as many as possible to observe physics and balance in the game, as observed in the quotes below.

"With personas, you can design tests to appeal to each type of persona. For example, the hardcore gamer is going to skip the tutorials and jump right in. While the noob will likely spend all the time in the introductory sections". (IN014)

"Have your system designers write edge case test plans for testers. They should also have an idea of where the system interacts with others." (IN032)

"The testers should be exploring edge cases and pushing the game to its logical limits, not validating your own wobbly code you couldn't be bothered to test." (IN038)

Finally, regarding testing automation on games, the findings confirmed the considerable high level of difficult to automate tests in this context. However, some evidence of successful cases of automation in games were identified and this finding can be suggestive to guide practitioners. Basically, there are parts of a game where test automation is impracticable, however, aspects related to game configuration, customizations, and main flow (optimal flow) can be automated in the direction of exhaustive testing, and also to reduce test time and effort. Therefore, the combination between automation and manual testing would be the best way to guarantee quality of a game, as illustrated below.

*"Also the freedom of movement present in most games and the randomization of other elements on a typical game that makes to feel it as more "realistic" are usually a nightmare for applying pure automated tests: in fact a human tester will find more quickly and more errors than any automated test. (IN022)*

*"Basic tests during our automated build process were a huge win. This included tasks such as creating a character, transferring maps, running some scriptable UI tests and looking for expected behavior." (IN063)*

*"Usually that involves automating test cases, but as games are graphics intensive, you can't automate screens and animations as tests, but your game logic could be automated." (IN066)*

In summary, the results of the case study were the first step to enlighten the particularities related to software testing in game development. However, there is no identified evidence regarding two specific phases of software-testing life cycle, namely, Test Execution and Test Closure. Therefore, more data was needed to be collected in order to identify as precise as possible the peculiarities of game testing and build a more complete body of knowledge about this theme. Therefore, we applied a survey questionnaire in a sample of practitioners in order to gather further evidence to this research.

## 4.2 Survey Research

This section starts presenting a brief description of the sample of individuals that participated of this survey, and then presents the summary description of the answers to the survey questions.

### 4.2.1 General Characterization of the Sample

The survey received answers from an amount of 41 professionals. Nevertheless, 3 individuals had no previous or current experience with game development or declared that they interact with games in only rare opportunities, therefore, they were excluded from the survey and a total of 38 individuals composed the final sample, which presents the following characteristics:

- Regarding geographic distribution, the final survey included individuals from 13 different countries, being 50% (10/38) of participants from Brazil, followed by 7.9% (3/38) from Canada and 7.9% (3/38) from Singapore. There was 2 participants from Germany and 2 participants from Finland, thus each country represents 5.9% (2/38) of the sample. Finally, there was 1 participant (2.6%) from each of the following countries in the sample: Romania, United Arab Emirates, Portugal, Norway, Sri Lanka, Morocco, Austria, Egypt and India;
- Regarding the current role or position, 31.6% of individuals (12/38) were working as software tester or testing activities, 31.6% of individuals (12/38) were software developers, 13.2% of individuals (5/38) were working with software design, 5.3% (2/38) were software requirements analysts and 5.3% (2/38) were software managers. Further, there were 13.2% of participants (5/38) not currently working with software development, but in academic position as researchers or professors. As these individuals had a background related to game development, they were included in the final sample;
- The average experience of the individuals in the sample is 5.5 years, in which the most experienced individual is working in software development for 30 years and the less experienced for less the one year. The standard derivation in this case was 6.3 years. Further, regarding experience in the current position, the sample presented an average of 2.41 years and standard derivation of 3.70 years.
- Regarding experience with games, when asked how often the individuals interact with games, over 47% (18/38) of participants answered *almost always,* while over 26% (10/38) of individuals answered *often* and over 26% (10/38) answered *sometimes*. This is an important information because some participants that answered the questionnaire affirmed that they *rarely* interact or have interacted with games, therefore, they were excluded from the final sample, in order to maintain the strength of the collected evidence.
- About testing automation, over 83% of the testers in the sample (10/12) affirmed that they have experience in automation and related activities in software testing. Further, other professionals such as developers affirmed they have work with software automation in some level.

After characterizing the survey sample, the answers for questions related to testing games were analyzed using the same qualitative process applied in the case study and described in Figures 2 and 3.

### 4.2.2 General Differences Between Software Testing and Game Testing

Practitioners pointed out the differences that they observed while working with games in comparison with their experience in working with general software development, and regarding software testing the differences can be grouped in three groups: user orientation, scope definitions and graphic details. All evidence collected in this phase of the survey is consistent with what was previously observed in the case studies.

Over 47% of individuals in the sample (18/38) believe that the main difference between testing a game and a regular software lies on the game requirements and the scope definitions of the project. Since both, games and regular software, have different aspects involved, expected results and variety of target users, requirements can be very unstable and scope definitions might be difficult to state, which will directly impact testing definitions, plan and estimations. This perception is illustrated in the quotes bellow.

*"I believe that one of the difficulties to test a game is the frequent changes this type of software can have over the time." (IN091)*

*"Even simple games could have complex rules and dynamic processes involved, everything is related to the scope of the project." (IN099)*

*"Games are not real things, hence you have to think what the gamer needs". (IN086)*

*"Testers need to know what is the main motivation for users to play and keep playing the game and then verify this." (IN094)*

Following this, over 42% of individuals believe that is more difficult to test a game than a regular software because of a range of different and specific human-centered interactions that might be presented in games. In this case, it would be relatively difficult to check the level of entertainment in games, depending on various types of players and their unpredictability while playing the game, as illustrated below.

*"I think games are more difficult to test than regular software, because a testing tool cannot measure the level of entertainment of the user." (IN077)*

*"Many software will work like a calculator, it's math, it will work or not! Game is like an experience, so it has different outcomes depending on the user, then you should considerate testing the human aspects behind the software." (IN097)*

A small percentage of participants (5% – 2/38) commented that the main particularity of games in comparison to regular software is related to graphical interface details, which is an extremely important element in games, however, is less representative in other types of software. Thus, since games are part of a very specific context, part of the professionals working in software testing might have limited experience regarding tests such graphical elements.

*"We should not only test the functional requirements of the game. Many games are considered "heavy" and consume a lot computational [graphical] resources." (IN079)*

*"Usually, in regular software, the main concern is whether the functionality is working or not, graphical details is something that demands more attention in games. So for games you have to separately test the game functionality and then, the performance, optimization and quality of graphics and images." (IN092)*

Finally, 5% of participants (2/38) affirmed that there are differences between testing games and testing a regular software. However, they did not provide a detailed opinion that could be analyzed.

*4.2.3 Particularities of Software-Testing Life Cycle in Games*

When asked about each individual phase of software-testing life cycle in games, practitioners confirmed the information gathered in the case study, regarding Requirements Analysis, Test Planning and Test Development. However, the survey collected information on how to overcome problems in these phases. Further, evidence was collected to characterize Test Execution and Test Closure, presenting the particularities of these two activities in the context of games.

In general, Requirement Analysis in game testing is characterized by practitioners as volatile and mutable, as previously observed. For this, simple practices could be effectively applied in order to improve this process in testing, such as effective communication among testers and all parties involved in the process (including users) and documented detailed information about the variety of users and their motivations to play the game.

*"Volatility and lack of expressiveness of game requirements are main challenges in testing games." (IN091)*

*"Communication between all parties involved. There is no way to gather all requirements at the beginning. Agile is better method." (IN085)*

*"Since games are from a context that is highly dependent on the state of art, everything needs to be carefully defined from the start, otherwise, changes can negatively impact test plans". (IN0111)*

Regarding, Test Planning and Test Case Development, there was no major differences between what was observed in the case study and what was collected on the survey. The data presented in the two studies is consistent, and no new evidence was identified. On the other hand, the survey was effective in collecting information about Test Execution, since no evidence about this phase of software-testing life cycle was identified so far. Thus, over 60% (23/38) participants agreed that there are specific types of bugs that are more common in games, and commented their experience about this question. Following this, the most common type of defect in games would be those related to physics and game rules, and since these are elements that directly affect user experience and aspects of entertainment and fun, testers need to be careful in verifying this kind of issues, as demonstrated below.

*"Yes, certainly, games provide more possibility of actions that were not planned then regular software, thus, the variety of rules open a variety of possibilities for bugs" (IN098)*

*"Especially the physics of the game and the mechanics of characters are more likely to crash." (IN104)*

*"Bugs that allow cheating or give advantages in multiplayer games are very uncommon in other types of software." (IN080)*

The second type of bug that is more common in games is related to images and the disposition and harmony of graphical elements on the screen and testers should be aware not just about this issues but also about related bugs, such as incompatibility problems, as follows.

*"Did you ever see online videos about bugs or crashes on Power Point? Of course not, but there is a series of videos showing bugs about visual deficiencies in games." (IN097)*

*"Yes, bugs are usually related to game appearance and aesthetics." (IN104)*

Practitioners also pointed out performance issues as something that needs attention from software testers, since the way that the game runs can provide good experience for players, especially when considering online games.

*"Yes, performance. It is important to check system delays." (IN078)*

*"There are failures related to screen refreshing rates and data transference that are critical for online games, for example." (IN104)*

Finally, regarding Test Execution, over 23% (9/38) of participants (9/38) believed that there is no specific type of bugs for games and over 15% of individuals (6/38) say they have no answer for this question or preferred not to opine.

The survey was also effective in identifying evidence about Test Closure in the software-testing life cycle considering the context of games, since no information regarding this activity was found in the case study. Despite of almost 58% of participants (22/38) did not believe that there are no particularities associated to this activity in game testing, the remaining of the sample believed that the final report in this context should include information about the coverage of the tests divided into at least three categories: coverage of requirements, coverage of flows and coverage of graphic elements. Thus, this report can be applied to improve testing plan activities in the future, regarding estimations of time, costs and allocation of professionals, depending on the type of game under development.

## 5 DISCUSSIONS

By applying a mixed-method approach based on a case study and a survey research, this study contributes to the improvement of the current body of knowledge regarding software testing and game development. During the development of this study, the evidence found in the literature was limited to a generic discussion on the limitation of test automation in games due to human-centered activities and human behavior presented in this context, and pointed out the lack of material and tools to automate. Further, researchers discussed on how friendly to changes games can be and hypothesized that the complexity and variant characteristics of different types of games would drive practitioners to a different process of reporting a failure, different from other types of software.

On the other hand, this current research gather a more extensive set of evidence presenting a more comprehensive group of information with general results regarding differences between software testing and game testing, and more importantly, specific particularities observable in each individual phase of software-testing life cycle. These specific aspects are summarized below.

**Table 2: Particularities of Game Testing**

| Test Phase | Observable Particularities |
|---|---|
| Requirements Analysis | 1. Game requirements are more susceptible to changes than regular software.<br>2. Changes in requirements are resultant of the exploratory process to better represent human-centered aspects on games.<br>3. Effective communication among test engineers and stakeholders involved in the process, especially game players, would be the best strategy to reduce the impact of requirements changing along the project.<br>4. In order to improve testing activities, game requirement specification might include detailed information about the different type of users targeted by the game, along with their motivations to play and to keep playing the game. |
| Test Planning | 5. Games are a type of software that often include uncommon measures that are difficult to access and therefore difficult to test, such as entertainment, fun and behavior.<br>6. Test plans should consider the complexity behind these measures in terms of estimation, schedule and resources allocation.<br>7. A strategy combining multiple approaches of unit tests, manual testing, automated tests and exploratory tests might be one of the keys to overcome the complex measure problem and guarantee coverage.<br>8. Game testing is a process in which the professional skills of testers can determine the level of success of the result. Therefore, test plans should include the allocation of testers that have familiarity with games. |
| Test Case Development | 8. Test cases creation is more complex in games than regular software, due to the enormous number of flows, input and outputs that could be involved in a single game action, which might be increased by number of different types of users.<br>9. The definition of people and the process of creating groups of players is a strategy that could help the creation of test cases, reducing the number of possible flows and enabling prioritization and coverage.<br>10. Due to singularity of some game requirements, automation scripts are very difficult to implement and maintain.<br>11. Some aspects of human interaction in games are impracticable for test automation. However, features related to game configuration, customizations and standard flows might be automated, reducing time and effort on the tests. Further, the groups of players and personas defined in the previous phases of the test process will improve the range of flows that could be automated. |
| Test Execution | 12. Testers must be aware of the most common types of bugs in games. In this type of software, issues could be concentrated in the game physics and rules, especially when there are multiple procedures to perform the same action. In addition, graphical elements and system performance deserve a careful checking process. |
| Test Closure | 13. Final test report and lessons learned might include details of tests coverage regarding requirements, flows and scenarios, and graphic elements. This information would be useful for feedback and planning future processes. |

We believe that the characterization of particularities in computer game testing presented in this study might improve the way games are developed and tested. Therefore, we expect that practitioners can successfully use the information summarized in this paper towards the improvement of games' quality. This is the main implication of this research.

Regarding threats to validity, we believe that the mixed-method approach applied in this research was effective to obtain diverse information and opinions from a variety of contexts, since the information gathered in the survey and the case study was collected from different types of professionals distributed over the world, working in different companies and with different projects. However, it is important to highlight that the participants on this study have different backgrounds, and despite of these professionals provide opinions based on their experience working with games, only a small percentage of individuals can be characterized as computer game testers. Further, considering validity, the consistency of the survey questionnaire was accessed through a pilot instrument with specialists from the field. Moreover, the data presented in the case study was collected by applying an exhaustive process verifying all the messages posted on online communities and selecting all those that were applied in the context of this research, using a well-defined process based on tested guidelines developed in the context of software engineering. Thus, we believe that the data collected in this study demonstrated good consistency.

## 6 CONCLUSIONS

Evidences about general differences between testing a regular software and a computer game, together with the observable particularities in the activities of the software-testing life cycle were identified as we applied the mixed-method based on a case study and an opinion survey, summarized using qualitative analysis techniques discussed in this paper. In summary, opinions based on the experience of 111 individuals were used to discuss this theme, contributing with the improvement of software testing in this context and expecting that the raised knowledge will help to increase the quality of computer games.

In general, games differ from regular software due to specific traits related to the complexity of human interactions characteristic in this type of software. Therefore, there are metrics that are difficult to observe and test, such as user behaviors, entertainment and fun, and these aspects might directly impact test activities, such test planning, development of test cases and even test execution. Thus, practitioners should be aware of these differences in order to improve the testing process, and this study is a step forward to this understanding.

## 7 ACKNOWLEDGMENT

Fabio Q. B. da Silva holds a research grant from CNPq #314523/2009-0. Cleyton V. C. Magalhães and Ronnie E. S. Santos are PhD students and receive a scholarship from CNPq. We are also very grateful to all participants for dedicating their time and attention to our research.